\documentclass{llncs}

\usepackage{latexsym}
\usepackage{graphicx,ulem,algorithm2e,algorithmic}
\usepackage {todonotes}
\usepackage{amsmath}

\newenvironment{keywords}{
       \list{}{\advance\topsep by0.35cm\relax\small
       \leftmargin=1cm
       \labelwidth=0.35cm
       \listparindent=0.35cm
       \itemindent\listparindent
       \rightmargin\leftmargin}\item[\hskip\labelsep
                                     \bfseries Keywords:]}
     {\endlist}
\usepackage{verbatim}

	\title{PROJECTION Algorithm for Motif Finding on GPUs}
	\author{Jhoirene B. Clemente, Francis George C. Cabarle, \and Henry N. Adorna}
	\institute{Algorithms \& Complexity Lab\\
Department of Computer Science\\
University of the Philippines Diliman\\
sDiliman 1101 Quezon City, Philippines\\
E-mail: \tt{\{jbclemente, fccabarle, hnadorna\}@up.edu.ph } 
}

\begin{document}	
\maketitle

\begin{abstract}
Motif finding is one of the NP-complete problems in Computational Biology. Existing nondeterministic algorithms for motif finding do not guarantee the global optimality of results and are sensitive to initial parameters. To address this problem, the PROJECTION algorithm provides a good initial estimate that can be  further refined using local optimization algorithms such as EM, MEME or Gibbs. For large enough input (600-1000 $bp$ per sequence) or for challenging motif finding problems, the PROJECTION algorithm may run  in an inordinate amount of time. In this paper we present a parallel implementation of the PROJECTION algorithm in  Graphics Processing Units (GPUs) using CUDA. We also list down several major issues we have encountered including  performing  space optimizations because of the GPU's space limitations.
\end{abstract}
		
\section{Introduction}\label{intro}

Motif finding problem (MFP) is the detection of overrepresented pattern  in sequences, and has become  a central problem in Computational Biology. Motifs can be transcription binding sites in DNA. These transcription sites regulate the expression of genes that are involved in similar cellular functions. Aside from identifying co-expressed genes,  motif finding can provide powerful hypotheses about links in the genetic regulatory networks \cite{manson}. Additionally, discovery of such patterns will help in the development of treatments and the identification of disease susceptibility \cite{shashidhara}.

Finding motifs is considered to be computationally hard, since it is considered to be an \textit{NP-complete} problem \cite{pevznerBook}. Existing algorithms are classified into two main categories: deterministic and nondeterministic\cite{yu}. Deterministic algorithms include the naive algorithm\cite{pevznerBook}, \textit{Statistical Enumerative Methods}, and \textit{Suffix Trees} \cite{pevznerBook}. Nondeterministic algorithms include \textit{Gibbs Sampling}\cite{yu}, \textit{Expectation Maximization}\cite{lawrenceEM} and \textit{Random Projection}\cite{tompa}. Deterministic algorithms assure the optimality of results,  however they are time consuming and are less effective for longer motifs. Nondeterministic algorithms are usually preferred for large input data,  since they do not require high computational complexity and are also effective for longer motifs. Nondeterministic algorithms do not guarantee the optimality of result since they are also sensitive to initial configurations. Aside from these two main categories, several algorithms use both techniques to compromise the advantages and disadvantages of using one of the two categories described. These hybrid algorithms include the works of \cite{shashidhara}, and Hybrid Gibbs sampling by \cite{hybridGibbs}.

With the flood of information and the volume of data we have nowadays, parallel algorithms and their implementation remain as important computational challenges in order to further boost performance and to keep up with the huge amount of data. Aside from traditional CPU-based grids and clusters the computational sciences, including biology, are using massively parallel hardware such as Graphics Processing Units (GPUs). GPUs are highly scalable (run thousands of cores easily), cheaper (by cost acquisition and maintenance) and provide performance increase and ease of use. Numerous Bioinformatics algorithms have been implemented using GPUs: BLAST, Smith-Waterman, Multiple Sequence Alignment, and Sequencing. Specifically, for motif finding there is Parallel Gibbs Sampling, CUDA-MEME\cite{liu}, and mCUDA-MEME\cite{liu}.  Brief discussions of these algorithms are presented in Section \ref{mfp_algorithms}.

\section{Motif Finding Problem}\label{motifFindingProblem}
\noindent  \textit{Regulatory motifs} in DNA are short sequences of patterns that occur in several locations of the genome as transcription binding sites or as a response to certain conditions. An example is the regulatory motif  in  fruit flies. Whenever flies are infected by pathogens (e.g. viruses, bacteria), they have an \textit{immunity gene} that is switched on to produce a certain protein used as an immune response. The gene responsible for the protein transcription will occur frequently at random positions on its genome. Finding motifs is not restricted to DNA, since it was first used in proteins for discovering transcription factors. Computationally, we can represent the genome and the patterns as strings, and the problem will be translated into finding a pattern of a given length that occurs frequently in the set of all strings. 
We first define notations before formally defining MFP.

\begin{itemize}

\item \textit{Sequences}$= \{S_1, S_2, ..., S_t\}$ denotes the list of $t$ strings, where each $S_i$ has length $n_i$. 

\item The alphabet $\Sigma$ is defined to be $\Sigma=\{A, C, T, G\}$ for DNA with cardinality $|\Sigma|=4$ and $\Sigma= \{A, R, N, G, C....\}$ for protein with cardinality $|\Sigma|=20$. 

\item The length of  a motif may be a range denoted by an ordered pair ($l^-$,$l^+$) or a constant $l$, where $l$, $l^{-}$, and $l^{+}$ are positive integers usually in the range $(5,20)$.

\item An $l$-mer is a string of length $l$ defined over some $\Sigma^{*}$. 

\item The notation $S_{ij}$ is an $l$-mer from $S_i$, starting with the $j$th position of the sequence.

\item The starting position vector $s=(a_1,a_2, ... a_t)$, contains the list of starting positions of each $l$-mer for each sequence, $1 \leq a_i \leq (n- l + 1)$.

\item An \textit{alignment matrix} $A(s)$ with dimension equal to $(t \times l)$ is derived from a vector of starting positions $s$, each row $i$ corresponds to an $l$-mer $S_{ia_i}$.

\item A \textit{profile matrix} $P(s)$ with dimension equal to $(|\Sigma| \times l)$ is derived from the frequency of each letter in each column of the \textit{alignment matrix}.
\end{itemize}

{$\{A, C, T, G\}$ represents the four nucleotide bases of the DNA: $Adenine$, $Cytosine$, $Thymine$, and $Guanine$ }respectively. {$\{A, R, N, G, C....\}$ are amino acid bases of proteins. 
Figure \ref{Sequences} gives an illustration of an instance of \textit{Sequences} while Figure \ref{alignment-profile} shows the corresponding $A(s)$ and $P(s)$.

\begin{figure}[h]
\centering
\begin{tiny}
\begin{tabular}{l c c c c c c c c c c c c c c c c c c c c c c c c c c c c c c c c c c c c c c c c}
$S_1:$ & C& G& G& G& G& C& T& A& T& G& G& A& A& C& T& G& G& G& T& C& G& T& C& A& C& A& T& T& C& C& C& C& T& T& T& C& G& A& T& A\\
$S_2:$ & T& T& T& G& A& G& G& G& T& G& C& C& C& A& A& T& A& A& A& T& G& C& C& A& C& T& C& C& A& A& A& G& C& G& G& A& C& A& A& A\\
$S_3:$ & G& G& A& T& G& C& A& A& C& T& G& A& T& G& C& C& G& T& T& T& G& A& C& G& A& C& C& T& A& A& A& T& C& A& A& C& G& G& C& C\\
$S_4:$ & A& A& G& G& A& T& G& C& A& A& C& T& C& C& A& G& G& A& G& C& G& C& C& T& T& T& G& C& T& G& G& T& T& C& T& A& C& C& T& G\\
$S_5:$ & A& A& T& T& T& T& C& T& A& A& A& A& A& G& A& T& T& A& T& A& A& T& G& T& C& G& G& T& C& C& A& T& G& C& A& A& C& T& T& C\\
$S_6:$ & C& T& G& C& T& G& T& A& C& A& A& C& T& G& A& G& A& T& C& A& T& G& C& T& G& C& A& T& G& C& A& A& C& T& T& T& C& A& A& C\\
$S_7:$ & T& A& C& A& T& G& A& T& C& T& T& T& T& G& A& T& G& C& A& A& C& G& T& G& G& A& T& G& A& G& G& G& A& A& T& G& A& T& G& C\\
\end{tabular}
\end{tiny}

\caption{$Sequences$ $=(S_1, S_2, S_3,`... , S_7)$, with $n_i= 40$ for $1 \leq i \leq 7$}
\label{Sequences}
\end{figure}


\begin{figure}[h]
\begin{center}
\begin{small}

\begin{tabular}{l l l c c c c c c c c}
					&$S_{1,8}$: 	&	A&T&G&G&A&A&C&T\\
					&$S_{2,19}$:	&	A&T&G&C&C&A&C&T\\
					&$S_{3,3}$:	&	A&T&G&C&A&A&C&T\\
\textbf{Alignment}	&	$S_{4,5}$: &	A&T&G&C&A&A&C&T\\
					&	$S_{5,31}$: &	A&T&G&C&A&A&C&T\\
					&$S_{6,27}$: &	A&T&G&C&A&A&C&T\\
					&	$S_{7,15}$: &	A&T&G&C&A&A&C&G\\
					\hline \\
					& \textbf{A}	: &	\textbf{7}&0&0&0&\textbf{6}&\textbf{7}&0&0 \\
					& \textbf{T}	: &	0&\textbf{7}&0&0&0&0&0&\textbf{6}\\
\textbf{Profile}	& \textbf{C}	: &	0&0&0&\textbf{6}&1&0&\textbf{7}&0\\
					& \textbf{G}	: &	0&0&\textbf{7}&1&0&0&0&1\\
					\hline \\
\textit{Consensus String}		&& \textbf{A} & \textbf{T} & \textbf{G} & \textbf{C}& \textbf{A} & \textbf{A}& \textbf{C}&\textbf{T} \\
					
\end{tabular}
\end{small}
\caption{the alignment  $A(s)$ corresponding to the set of strings $\{S_{1,8}, S_{2,19}, S_{3,3} ,S_{4,5}, S_{5,31}, S_{6,27}, S_{7,15})$ and the  profile matrix $P(S)$ with $l=8$, and $s=(8,19,3,5,31,27,15)$}
\label{alignment-profile}
\end{center}
\end{figure}

From the $P(s)$ we define $M_{P(s)}(i)$, where $1\leq i\leq l$, be the maximum number at $i$th column of the profile matrix (elements of  $P(s)$ written in boldface) e.g. $M_{P(s)}(1)$ = 7, and $M_{P(s)}(5)$ = 6 .  We define a \textit{consensus string} to be an $l$-mer, where each of its element is the nucleotide base corresponding to $M_{P(s)}(i)$. For instance, the consensus string in Figure \ref{alignment-profile} is `ATGCAACT', and there can exist an instance where $M_{P(s)}(i)$ will correspond to more than one letter so we choose one arbitrarily from the choices. Thus the consensus string for a given $P(s)$ is not unique. We define the \textit{Score(s,DNA)}, to be equal to $$\sum^{l}_{i=1} M_{P(s)}(i).$$ The \textit{Score} in the above example is $7+7+7+6+6+7+7+6= 53$. We use the \textit{Score} to evaluate the strength of a given alignment. Below is the formal definition of the MFP on DNAs.

\begin{definition} \label{mfpDefinition} \textbf{Motif finding Problem(MFP)} \\
\noindent INPUT: A $(t \times n)$ matrix of DNA, and $l$, the length of the pattern to find\\
OUTPUT: An array $s$= ($a_1, a_2, . . . , a_t$) maximizing \textit{Score(s,DNA)}
\end{definition}

MFP maximizes the \textit{Score(s,DNA)} function over all possible starting positions. The maximum value for \textit{Score} is $l \cdot t$ which corresponds to the best possible alignment, where each row for each column has the same element. However, having a consensus score equal to $\frac{l \cdot t}{4}$ corresponds to the worst possible alignment, where each column has an equal distribution of nucleotide bases. The algorithm for the naive MFP is given in Algorithm \ref{naivemfp}. 

\begin{algorithm}\label{naivemfp}
\KwIn{$DNA$, $l$}
\KwOut{$bestMotif$}
~\\
\begin{verbatim}    
    Procedure NAIVEMFP:
        bestScore <- 0
        For each s in (1,1,..1) to (n-l+1, n-l+1,...n-l+1)
        	if Score(s,DNA) > bestScore
    	       bestScore <- Score(s,DNA)
    	       bestMotif <- Consensus(s)
        return bestMotif
    end Procedure NAIVEMFP	
	
\end{verbatim}	
\caption{Algorithm for naive MFP solver.}
\end{algorithm}

Note that in Algorithm \ref{naivemfp} we need to exhaust all possible starting positions and compute the \textit{Score} to be able to determine $s$ with the best possible alignment. The total number of starting positions to test is $(n-l+1)^t$, which is exponential in $t$. The computation of \textit{Score(s,DNA)} will take $O(l)$. Therefore, the running time complexity of the naive MFP algorithm is $O((l \cdot n)^t)$.
To reduce the running time complexity of motif finding, \cite{pevznerBook} uses the concept of median string to exhaust the possible motifs. With the inclusion of this idea, the running time complexity of finding motif  is reduced to $O(|\Sigma|^l \cdot n \cdot t)$.

For these algorithms, the assumption is that, each sequence contains exactly one occurrence of the motif. This assumption follows the basic search model called \textit{OOPS} (One Occurrence per Sequence). Other models are \textit{ZOOPS} and \textit{TCM}.\textit{ ZOOPS} (Zero or One Occurrence per Sequence) is the generalized model for \textit{OOPS}, where we assume that for each sequence the occurrence of the motif is at most one i.e. we allow a sequence without any occurrence of the motif. \textit{TCM} (Two component mixture) model assumes that each sequence may contain zero or more occurrence of the motif. 

In some cases, occurrences of motif in sequences may vary due to mutations or genome rearrangements, and  motif finding algorithms become even more challenging. Because of this reason, \cite{pevznerBook} defined a specific variant of MFP in which we expect $d$ mutations from each occurrences of the motif. This variant is called $(l,d)-$motif finding, and is similar to our previous definition except for the fact that each occurrence(s) of motif $M$ in sequence $i$, $0<i<(t-1)$, the edit distance of $M$ in each of those occurrence(s) is $d$. 

Experiments from \cite{tompa} showed that their \textit{PROJECTION} algorithm performs better than existing algorithms in solving this specific variant. In addition to this, most of the challenging problem instances like (14-4), (16,5), and (18,6)  were successfully computed. Details for the PROJECTION algorithm will be discussed in Section \ref{projection}.

\section{Algorithms for Motif Finding}\label{mfp_algorithms}
Algorithms for motif finding can be classified into two main categories: deterministic (\textit{word-based enumerative methods}) and nondeterministic (\textit{probabilistic sequence models}). The \textit{word-based models}  rely on exhaustive enumeration, counting and comparing nucleotide frequencies. These algorithms include the naive implementation in Algorithm \ref{naivemfp}, branch and bound techniques, and suffix trees. Meanwhile, the \textit{probabilistic sequence models} rely on model parameters that are estimated using Likelihood computation and Bayesian inference. This category includes algorithms such as Gibbs Sampling\cite{yu}, Expectation Maximization\cite{liu} and Random Projection\cite{tompa}.

Word-based methods guarantee global optimality of result, and are appropriate for short motifs. However, for longer motifs and weakly constrained positions, word-based methods may become problematic due to their high computational complexity. The probabilistic sequence models require few search parameters, and are more effective for longer and general motifs. These algorithms are  dependent on probability matrices, and are called  Position Specific Scoring Matrix(PSSM), $\theta$ model, and frequency matrix in Gibbs Sampling, Expectation Maximization and Random Projection respectively. The probabilistic sequence models however do not guarantee the global optimality of the results, but these models do run significantly faster than the previous models by giving good enough solutions.

\section{GPU computing}

In 2007 NVIDIA, a well-known graphics processor manufacturer, introduced the \textit{Compute Unified Device Architecture (CUDA)} as an update of their G80 line of GPUs \cite{cudabook}.
CUDA provides numerous advances in massively parallel computations: (a) allowing programmers to declare only the resources they need (threads/ computatonal units and memory space used). Though additional conversion is needed for the input data, CUDA can significantly reduce the running time of methods related to string comparison, among other methods (b) hardware improvements such as abstracted performance scaling from a GPU with fewer cores to another with more cores (c) introduction of memory and computational unit hierarchy, (d) latency hiding etc. \cite{cudaguide,cudabook}. All of these help boost parallel computations in GPUs compared to CPU-based clusters or grids which require enormous amounts of money to setup and maintain. Comparable CPU cluster performance can be achieved by buying graphics cards from computer resellers at about \$500 USD containing up to 512 scalar processors at the moment. In order to maximize the significant speedups of GPUs, the computation should be expressed as much as possible in a data-parallel form, mapping it into the massively parallel GPU. Otherwise computations may not achieve significant speedups over CPU computations.
\begin{figure}[h]
\begin{small}
\centering
\includegraphics[scale=.35]{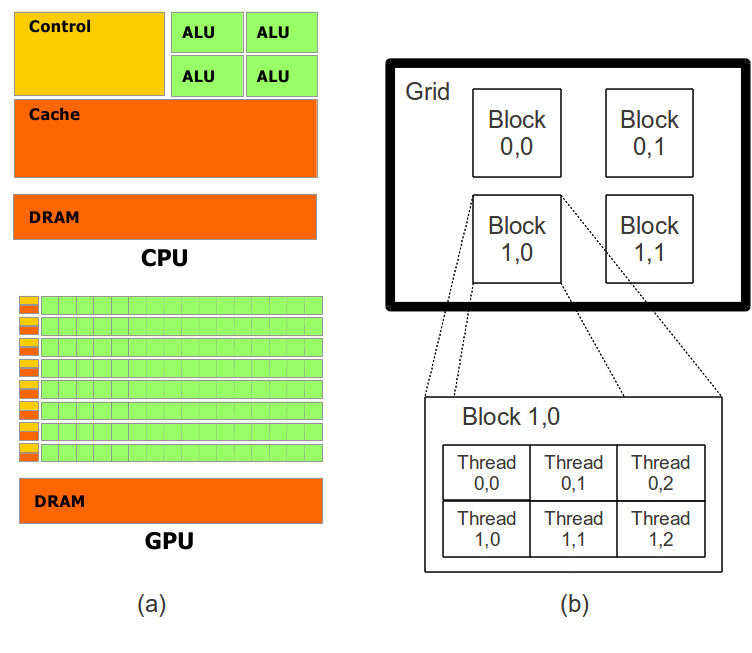}
\caption{(a) Performance increase of GPUs over CPUs are due to the difference in their design philpsophies: CPUs usually devote transistors largely to caching, control, DRAM, and the few left to arithmetic and logic, while GPUs devote most of their transistors to arithmetic and logic, with some on DRAM and few left to caching and control \cite{cudabook,gpgpu} (b) Computational unit hierarchy of CUDA, showing the grid, thread blocks, and threads. } 
\label{gpu-vs-cpu}
\end{small}
\end{figure}

A CUDA enabled GPU consists of, from largest to smallest, a grid of thread blocks, a block of threads, and individual threads \cite{cudaguide}. Threads are the atomic computational units in the GPU, but the programmer can specificy how to organize the grid, blocks, and threads based on their needs. \textit{CUDA C}, an extension of the \textit{C} language, provides functions to manipulate the GPU. A \textit{kernel function} is one that will run in the \textit{device}, another term synonymous to GPUs in the context of GPU programming. The kernel function is declared in the \textit{host} (synonymous to the CPU/s). The grid in the GPU can have a maximum grid dimension values of $(65535 \times 65535)$ thread blocks (i.e. it is 2-dimensional) while each block can have a dimension of $(512 \times 512 \times 64)$ threads per block (i.e. 3-dimensions), and all of these operate in parallel. With these resources available it is easy to see how massively parallel computations can be performed. 

\section{PROJECTION Algorithm}\label{projection}

Random projection  is a statistical technique commonly used for dimensionality reduction and visualization. Given a vector $v$ with $n$ dimensions, a projection $h_k(v)$ is a $k$ dimensional vector from $v$,  where elements in $h_k(v)$ are randomly chosen from $v$ and the order of elements are preserved. The PROJECTION algorithm discussed by \cite{tompa} uses this technique to solve motif finding and specifically to a challenging variant called $(l,d)$-motif finding and is showin in Algorithm \ref{projectionalgo}.

\begin{algorithm}\label{projectionalgo}
\noindent 
\noindent\textbf{INPUT:} $t$ sequences, motif length $l$, sequence length $n$, pairwise edit distance of all motif occurrences $d$\\
\textbf{OUTPUT:} \textit{Motif}
\begin{enumerate}
\item If not specified, determine the optimal value for $k$,  bucket threshold $s$, and  number of independent trials $m$.
\item For each independent trials, 
\begin{enumerate}
	\item Generate $(l-k)$ random numbers in range $[0,l]$.
	\item For each l-mer $x$ in $t$ sequences, get $h_k(x).$	
	\item Hash each $k$-mer$(h_k(x))$ to $|\Sigma|^k$ bins.
\end{enumerate}

\item Identify enriched bucket(s).
\end{enumerate}
\caption{PROJECTION algorithm}
\end{algorithm}

There are three main parameters in Algorithm \ref{projectionalgo}, the projection value $k$, the bucket threshold $s$, and the number of independent trials $m$. The identification of the optimal value of these three main parameters are discussed in \cite{tompa}.

For the $(l,d)$-motif problem, if we have $M$ as the planted motif, and each occurrence $M_i$ in $S_i$ has an edit distance $d(M,M_i)=d$. The projection dimension $k$ should be at least the number of matched strings in $M$ and $M_i$ i.e. $l-d$, to lessen the probability that the mutated letters are hashed. We would also want $k$ to be as high as possible to capture the original motif. According to \cite{tompa}, the optimal value for $k$ is $(l-d-1)$.   Also, to minimize the contamination of planted buckets by random background projection, an optimal value for bucket threshold $s$ should be twice the average of the bucket size $t\cdot(n-l+1)/4^k$. However, in the challenge problem instances from \cite{tompa}, $s$ is always negative because $t\cdot(n-l+1) << 4^k$. Based from their empirical testings, values $s = (3,4)$ for challenge problems output a significant motif. 

Increasing $m$ would give us a motif that is closer to the global optimal solution. To identify the optimal value of $m$, we need to specify a probability $q$ that the planted bucket contains $s$ or more instances in at least one of the $m$ trials. The computation for $m$ is 

$$m= \lceil{\frac{\log{(1-q)}}{\log{B_{\hat{t},\hat{p}(l,d,k)}}(s)}}\rceil, $$
where $\hat{t}$ is the estimate of the number of input sequences containing a planted motif, and $\hat{p}(l,d,k)$ is the probability that a given planted motif is hashed to the enriched bucket, which is equal to $$\hat{p}(l,d,k)= \frac{{(l-d) \choose k}}{{l \choose k}}.$$ In a given trial, the probability that fewer than $s$ $k$-mers are hashed in an enriched bucket is $B_{\hat{t},\hat{p}(l,d,k)}(s)$.

\subsection{Motif Refinement using Expectation Maximization (EM)}
\indent After identifying enriched bucket(s), \cite{tompa} performed motif refinement using EM. This algorithm involves two steps, getting the  $Expectation$ of a model (\textit{E-step}) and maximizing it (\textit{M-step}). The model with the highest $Expectation$ yields the most probable motif. Although EM is a local optimization algorithm, i.e. does not assure global optimality and is very sensitive to initial configuration, the initial motif model from PROJECTION significantly increases the chance of getting the optimal solution. To illustrate how  EM works, refinement  of one identified enriched bucket is shown below in Example \ref{exampleMotifModel} from the given $Sequences$ in Figure \ref{Sequences}.

\begin{example}\label{exampleMotifModel}
If positions $1,\ 2,\ 3,\ 5,$ and $6$ are used for projection and bucket threshold $s$ is equal to $4$.
We identified bucket \textit{``ATGAC"} to be enriched with $k$-mer count equal to $7$. The corresponding $l$-mers hashed in the bucket are $\{S_{1,8}, S_{2,19}, S_{3,3} ,S_{4,5}, S_{5,31}, S_{6,27}$, and $S_{7,15})$, which yield the alignment shown in Figure \ref{alignment-profile}. 

From the alignment, we define a motif model $\theta^0$ that serves as the initial parameter for EM. Motif model $\theta$ is a $(t\ \times\ (l+1)) $ matrix where each element in $\theta$ is equal to 
\begin{equation*}\small
\theta_{i,j} = 
\begin{cases}
\text{probability of $\sigma_i$ appearing at position $j$ of the motif,}  &  \text{if $1 \leq j \leq l$}\\
\text{probability of $\sigma_i$ appearing outside of the motif,} & \text{if $j=0$}
\end{cases} 
\end{equation*}
\noindent where $\sigma_i$ is an element of the the input alphabet $\Sigma = \{\sigma_1, \sigma_2, \ldots\}$. 
\begin{figure}[h]
\begin{center}
\begin{small}
\begin{tabular}{l l l c c c c c c c c}
					&$S_{1,8}$: 	&	A&T&G&G&A&A&C&T\\
					&$S_{2,19}$:	&	A&T&G&C&C&A&C&T\\
					&$S_{3,3}$:	&	A&T&G&C&A&A&C&T\\
\textbf{Alignment}	&	$S_{4,5}$: &	A&T&G&C&A&A&C&T\\
					&	$S_{5,31}$: &	A&T&G&C&A&A&C&T\\
					&$S_{6,27}$: &	A&T&G&C&A&A&C&T\\
					&	$S_{7,15}$: &	A&T&G&C&A&A&C&G\\
					\hline \\
					& \textbf{A}	: 0.25 &	\textbf{1.00}&0.00&0.00&0.00&\textbf{0.86}&\textbf{1.00}&0.00&0.00 \\
					& \textbf{T}	: 0.25 &	0.00&\textbf{1.00}&0.00&0.00&0.00&0.00&0.00&\textbf{0.86}\\
\textbf{$\theta^{0}$}	& \textbf{C}	: 0.25 &	0.00&0.00&0.00&\textbf{0.86}&0.14&0.00&\textbf{1.00}&0.00\\
					& \textbf{G}	: 0.25&	0.00&0.00&\textbf{1.00}&0.14&0.00&0.00&0.00&0.14\\					
\end{tabular}
\end{small}
\caption{Computation of initial motif model $\theta^0$ with $Expectation = 8.58 $, equal to the summation of maximum values per column, except for the first one. First column corresponds to the background probability of the the given sequence. For this example we generated the data such that each symbol has equal frequency in $S$}
\label{profile}
\end{center}
\end{figure}
In this particular example, the bucket \textit{``ATGAC"} is the planted bucket of  $Sequences$ shown in Figure \ref{Sequences}. The $Expectation$ computed is the maximum and therefore has already converged with a consensus string equal  to \textit{``ATGCAACT"}. However for real data, or challenge motif problems with $n$ equal to $(600-1000)$, several enriched buckets may contain random background $k$-mer and may lead to false motifs. This is the reason why identification of the three key parameters is crucial.
\end{example}
Other local optimization algorithms can be substituted for EM. Some hybrid algorithms  for motif finding use PROJECTION to provide an initial motif model, in some instances only a single iteration or trial is needed. 

\subsection{Analysis}
To improve the running time of an algorithm, space complexity is one of the trade-offs. We present two possible versions of PROJECTION algorithm. One with fast running time, with a very high space complexity (PROJECTION1), and another with lower space complexity and a slower running time (PROJECTION2). Let $x = t \cdot (n-l+1)$ be equal to the number of $l$-mers in $Sequences$.

In PROJECTION1, projecting all $l$-mers to $k$-mers need $O(l \cdot x)$ time with $O(l \cdot x)$ space. Hashing requires $O(x)$ time with $O(|\Sigma|^k)$ space. Getting the list of enriched buckets will take $O(x)$ with space $O(r \cdot x)$, where $r$ is the maximum number of $k$-mer hashed in a bucket.  Therefore, PROJECTION1 needs a total of $O(x)$ time. We do not consider $O(l \cdot x)$, because the value of $l$ is negligible compared to $x$ with space $O(|\Sigma|^k)$.

In PROJECTION2, projecting all $l$-mers to  $k$-mers need $O(l \cdot x)$ time with $O(l \cdot x)$ space. Hashing requires $O(x^2)$ time with $O(x)$ space. {Hashing is done without maintaining actual buckets, but each $k$-mer is compared to all other $k$-mers}. Getting the list of enriched buckets will take $O(x)$ with space $O(r \cdot x)$, where $r$ is the same from PROJECTION1.  Therefore, PROJECTION2 needs a total of $O(x^2)$ time with space $O(x)$.

\section{CUDA-PROJECTION}
PROJECTION algorithm is highly parallelizable. One can think of each thread in the GPU projecting and  hashing an $l$-mer  to its corresponding bin. However, maximizing  the resources available becomes a nontrivial task.  In the GPU, we can launch thousands of threads running in parallel so running time is usually compensated, but the amount of device (GPU) memory is more limited so space has to be compensated by the algorithm. The parallelization scheme we used is briefly discussed  in Algorithm \ref{cuda-projection}.

\begin{algorithm}\label{cuda-projection}
\textbf{INPUT:} $t$ sequences, motif length $l$, sequence length $n$, pairwise edit distance of all motif occurrences $d$\\
\textbf{OUTPUT:} \textit{Motif}
\begin{enumerate}
\item If not specified, determine the optimal value for $k$,  bucket threshold $s$, and  number of independent trials $m$.
\item For each independent trials, 
\begin{enumerate}
	\item Generate $(l-k)$ random numbers in range $[0,l]$.
	\item  For each l-mer $x$ in $t$ sequences, get $h_k(x)$	 \textit{in parallel.}
	\item Hash each $k$-mer$(h_k(x))$ to $|\Sigma|^k$ bins \textit{in parallel.}
\end{enumerate}

\item Identify enriched bucket(s)  \textit{in parallel}.
\end{enumerate}
\caption{CUDA-PROJECTION}
\end{algorithm}

For each independent trials in PROJECTION, we generate $(l-k)$ random numbers that range from $0$ to $l$. Given that the optimal value for $k$ is $(l-d-1)$, we know that $(l-k)$ is always less than $k$. Therefore, instead of generating $k$ random positions to be projected, we generate $(l-k)$ random positions not to be projected. Also, not parallelizing each independent trial leaves the task of generating unique random seeds. Random number generation is still done in host (CPU).

The parallelization starts after calling a kernel function that project and hash each $l$-mer in $t$ sequences. Before doing necessary computations, the host needs to transfer the input data and allocate space to the device memory.  To ease up the manipulation in GPU, instead of passing multidimensional arrays, we passed arrays in linear form.
\begin{figure}[h]
\centering
\includegraphics[scale=.4]{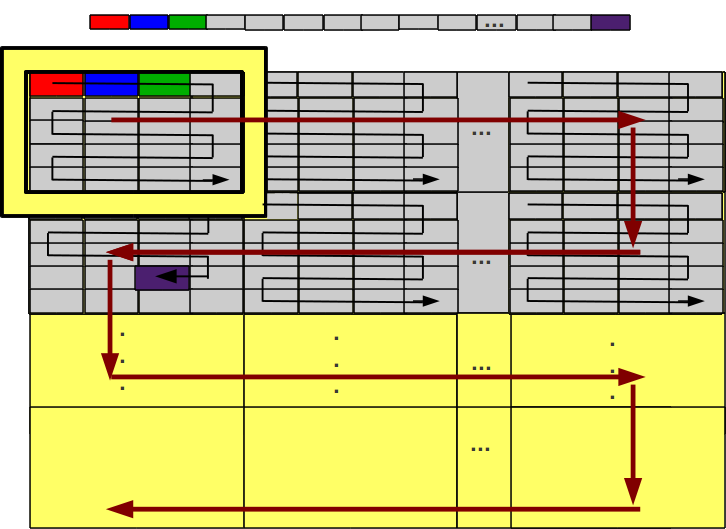}
\caption{Illustration of the logical and actual arrangement of threads in GPU for CUDA-PROJECTION. Each of the processor holds an $l$-mer in $Sequences$.   } 
\label{threads}
\end{figure}
 In the kernel function, each thread access each $l$-mer $x$ in the sequence list and compute $h_k(x)$ using the random numbers generated earlier.
\begin{figure}[h]
\centering
\includegraphics[scale=.5]{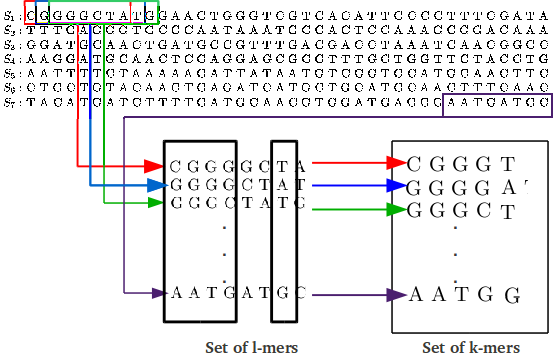}
\caption{Illustration of how each thread (represented by an arrow) accesses and projects each $l$-mer in $Sequences$ to produce a list of all $k$-mers. In this example, hashing function $h_k$ uses positions $(1,2,3,4, 7)$ in the projection.}
\label{project}
\end{figure}

For each  $k$-mer produced by projection, we computed an integer equivalent to a unique string. The computation follows the conversion of  base-$|\Sigma|$   to base-$10$. We defined a one-to-one mapping of integers $\{0, .. ,|\Sigma|\}$ to the set of alphabets $\Sigma$. For a DNA input, we defined a mapping $\{(A:0), (C:1), (T:2), (G:3)\}$. For example, string \textit{``AAAAAAAA"} is equivalent to integer $0$, and \textit{``AAAAAAACG"} is equal to $7$. The integer representation of each $k$-mer reduces the time complexity of string comparison.

Instead of hashing all $k$-mers to $4^k$ buckets which uses $O(4^k)$ space,  we implemented another method to simulate the hashing using $O(t \cdot t \cdot (n -l +1) )$ space.  Since this is done in parallel, each thread will compare a specific $k$-mer to all other $k$-mers.  Each thread will take note of the id of the threads whose $k$-mer matches its own $k$-mer. Each thread will have a list of thread IDs $pid(k-mer) = \{pid^{'}_0, ..., ,pid^{'}_c\}$, where 
\begin{center}
 $k$-mer($pid$)  $=$  $k$-mer($pid^{'}_0$) $=$ $\ldots $ $=$ $k$-mer($pid^{'}_c$),
 \end{center}
 and then evaluates whether it satisfies the bucket threshold (i.e. $c\geq s$). 

\begin{figure}[h]
\centering
\includegraphics[scale=0.3]{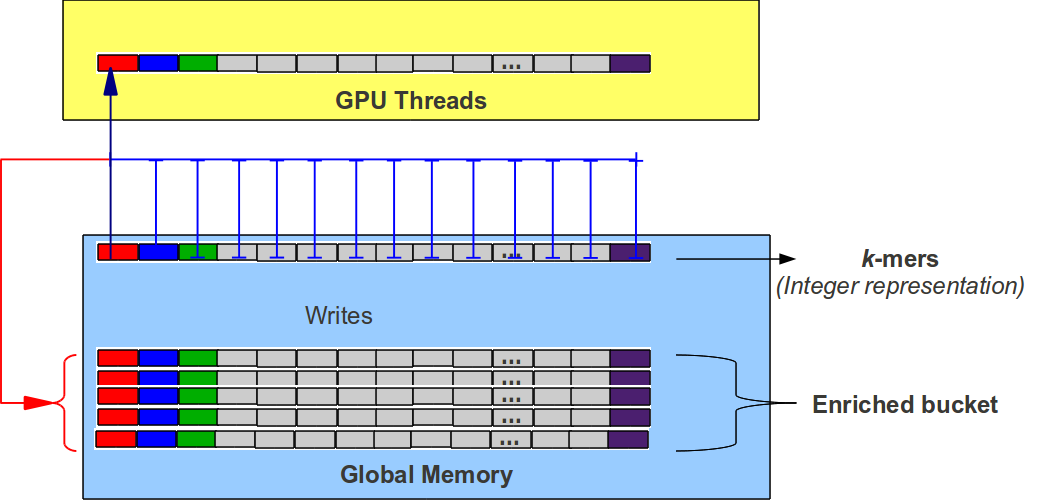}
\caption{Illustration of how each thread in the device compares one $k$-mer to all other $k$-mers (Blue arrows). Once a $k$-mer finds a match, it writes the thread id ($pid$) to the allocated space in enriched buckets array (Red arrow). }
\label{getEnriched}
\end{figure}
To illustrate how we get the list of enriched buckets, Figure \ref{getEnriched} shows how each thread reads from the list of $k$-mers  and writes to the  array of enriched buckets. Given the following scheme, not every bucket in the list of enriched buckets will continue to motif refinement, only those which satisfy the threshold value.  Given that we have a maximum of $b$ $k$-mers hashed in a bucket, we have at most $b$ buckets that would yield the same motif in a single trial. 

Each of the identified enriched bucket will proceed to motif refinement using EM. The algorithm will iterate in constant time,  as suggested by \cite{tompa}. Based from their empirical testings, EM usually converges before the 5th iteration. Given that one of the enriched bucket is the planted bucket, EM will return the planted motif. 
One trial of CUDA-PROJECTION does not guarantee the return of the planted motif.  Therefore, CUDA-PROJECTION requires $m$ iterations to  increase the probability of returning the planted bucket.

The CUDA-PROJECTION algorithm uses $x=  t \cdot (n - l +1) $ threads. Each thread hashes, projects and evaluates each bucket whether they are  enriched or not, in parallel. We allocated device memory spaces for  for the list of all sequences, $l$-mers, $k$-mers, buckets, and enriched buckets. The list of all sequences takes $O(t \cdot n)$ space. The list of all $l$-mers takes $O(l \cdot x)$ space.  The list of all $k$-mers takes $O(k \cdot x)$ space. The bucket  list takes  $O(x)$ space. The enriched bucket list takes $O( r \cdot x)$ space, where $r$ is the maximum number  of $k$-mers hashed in a bucket.

The enriched bucket contains the list of all thread IDs hashed with the same $k$-mer. In our implementation the constant  $r$ is  $(t\cdot s)$, because (1) we know that the number of hashed $k$-mers in a bucket should at least be equal to the bucket threshold $s$, (2)  we also know that the planted bucket should contain at least $t$ $k$-mers, given that all projected $k$-mers do not contain the mutated characters, and (3) we assume that  background $k$-mers may hash to the planted bucket. Although enriched buckets maintain space quadratic in $t$, compared to exponential space $4^k$, it is optimal. The comparison between the two space complexities is shown in Figure \ref{spaceComparison}	. The $y$-axis represents the size, and $x$-axis are independent challenge $(l,d)$ problems, sorted with respect to $l$. Note that for $ l > 17$, $4^k$ is always greater than  $(r \cdot x)$, given that $t=20$ and $n=600$.
 \begin{figure}
 \centering
 \includegraphics[scale=.60]{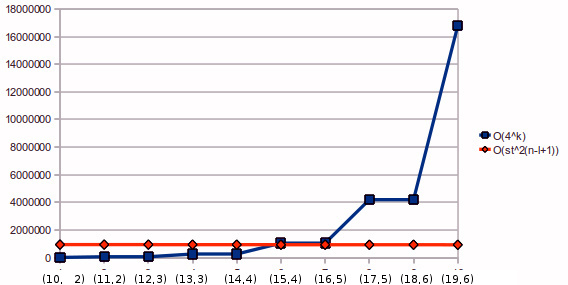}
\caption{Space complexity used compared to $4^k$ space} 
\label{spaceComparison}
 \end{figure}

In CUDA-PROJECTION, projecting all $l$-mers to $k$-mers need $O(l)$ time with $O(l \cdot x)$ space. Hashing requires $O(k)$ time ({the conversion of a string with length $k$  to an integer of base-$10$ needs $O(k)$}) with $O(x)$ space. The space complexity is {not a factor of $k$ because of the base-10 conversion of each $k$-mer}. Getting the list of enriched buckets will take $O(x)$ with space $O(r \cdot x)$, where $r$ is the maximum number of $k$-mer hashed in a bucket.  Therefore, CUDA-PROJECTION needs a total of $O(x)$ time with space $O(x)$.

\section{Conclusion}
We have presented a parallel implementation of the PROJECTION algorithm for motif finding on the GPU. The CUDA-PROJECTION implementation  runs in $O(x)$ time with space complexity $O(x)$. Compared to PROJECTION1, our implementation achieved no speedup but we significantly reduced the space complexity. Compared to PROJECTION2, our implementation uses the same amount of space with linear speedup. In other words, CUDA-PROJECTION captures the best side of both sequential implementations. 

\section*{Acknowledgments}
Jhoirene Clemente and Francis George Cabarle are supported by the {DOST-ERDT program}. Henry Adorna is funded by the {DOST-ERDT research grant} and the Alexan professorial chair of the {UP Diliman Department of Computer Science}.


\bibliographystyle{ieee}

\begin{thebibliography}{99}
\bibitem{bailey} Bailey T L., ``Discovering Motifs i DNA and Protein Sequences: The Approximate Common Substring Problem", 1995, Ph.D. Dissertation, University of California San Diego
\bibitem{tompa}Buhler J., and Tompa M., ``Finding Motifs Using Random Projections", 2001, RECOMB '01 Proceedings of the fifth annual international conference on Computational biology
\bibitem{chen}Chen C., Schmidt B., Liu W., and Muller-Wittig W., ``GPU-MEME: Using Graphics
Hardware to Accelerate Motif Finding in DNA Sequences.", 2008, LNCS 5265, 448-459.
\bibitem{dempster} Dempster A P., Laird N M., and Rubin D B., ``Maximum Likelihood from Incomplete Data via the EM Algorithm", 1977, Journal of the Royal Statistical Society, Series B
\bibitem{gpgpu} M. Harris, ``Mapping computational concepts to GPUs'', {\it ACM SIGGRAPH 2005 Courses}, NY, USA, 2005.
\bibitem{hertz} Hertz G Z., and Stormo G. D.,``Identifying DNA and Protein Patterns with Statistically Significant Alignments of Multiple Sequences", 1999, Bioinformatics, 15:563-577
\bibitem{pevznerBook} Jones N.,and Pevzner P.,``An Introduction to Bioinformatics Algorithms", 2004, Massachusetts Institute of Technology Press 
\bibitem{cudabook} D. Kirk, W. Hwu, {\it Programming Massively Parallel Processors: A Hands On Approach}, 1st ed. MA, USA: Morgan Kaufmann, 2010.
\bibitem{manson} McGuire A., and Church G., ``Discovery of DNA Regulatory Motifs", Harvard University Medical School
\bibitem{lawrenceEM} Lawrence C., Reilly A., ``An Expectation Maximization Algorithm for the Identification and Characterization of Common sites in Unaligned Biopolymer Sequences.", 1990, Proteins 7 (1), 41-51
\bibitem{lawrenceGibbs} Lawrence C., Altschul S., Boguski M., Liu J., Neuwald A.,and Wootton J., ``Detecting subtle sequence signals: A Gibbs sampling strategy fr multiple alignment", 1993, Science 262, 208-214 
\bibitem{liu} Liu Y., Schmidt B., Liu W.,and Maskell D. , ``CUDA-MEME: Accelerating Motif Discovery in Biological Sequences Using CUDA-enabled Graphics Processing Units",2009, Pattern Recognition Letters 31, 2170-2177
\bibitem{parkMiller} Park S., and Miller K.W.,``Random Number Generator: Good ones are Hard to Find", 1988, COMM, ACM 31, 1192-1201
\bibitem{shashidhara} Shashidhara H S., Joseph P. ,and Srinivasa K G., ``Improving Motif Refinement using Hybrid Expectation Maximization and Random Projection", ISB 2010, Feb 15-17, Calicut India
\bibitem{hybridGibbs} Shida K., ``Hybrid Gibbs-Sampling Algorithm for Challenging Motif Discovery: GibbsDST", 2006, Genome Informatics 17(2): 3-13
\bibitem{yu} Yu L., and Xu Y., ``Parallel Gibbs Sampling Algorithm for Motif Finding on GPU", 2009, IEEE International Symposium on Parallel and Distributed Processing with Applications
\bibitem{cudaguide} NVIDIA corporation,{\it ``NVIDIA CUDA C programming guide''}, version 3.2, CA, USA: NVIDIA, 2011.
\end{thebibliography}

\end{document}